\documentclass[twocolumn,
               superscriptaddress,
               floatfix,
               longbibliography, 
               showkeys,apl, nofootinbib]{revtex4-1}

\usepackage{graphicx} 
\usepackage{bm}	
\usepackage{color}                     
\usepackage{xcolor}
\usepackage{epsfig}
\usepackage{amsmath, bm} 
\usepackage{amssymb} 
\usepackage{longtable} 
\usepackage{float}
\usepackage{mathtools}
\usepackage{xparse}
\usepackage{hyperref}
\usepackage{times}
\usepackage[normalem]{ulem}
\usepackage{braket}
\usepackage{multirow}
\usepackage{textcomp}

\usepackage[ruled]{algorithm}

\usepackage{algpseudocode}
\usepackage{enumitem}
\usepackage[caption=false]{subfig}

\usepackage{sidecap}
\sidecaptionvpos{figure}{t}

\newcommand{\be}{\begin{equation}}
\newcommand{\ee}{\end{equation}}
\newcommand{\bd}{\begin{displaymath}}
\newcommand{\ed}{\end{displaymath}}
\newcommand{\BE}{\begin{eqnarray}}
\newcommand{\EE}{\end{eqnarray}}

\definecolor{darkgreen}{rgb}{0.0, 0.5, 0.0}

\begin{document}

\setlist[enumerate,1]{label=\arabic*, start=0}

\title{A Framework for Demonstrating Practical Quantum Advantage: Racing Quantum \\ against Classical Generative Models}

\author{Mohamed Hibat-Allah}
\affiliation{Zapata Computing Canada Inc., 25 Adelaide St East, M5C 3A1, Toronto, ON, Canada}
\affiliation{Vector Institute, MaRS Centre, Toronto, Ontario, Canada M5G 1M1}
\affiliation{Department of Physics and Astronomy, University of Waterloo, Waterloo, Ontario, Canada N2L 3G1}

\author{Marta Mauri}
\affiliation{Zapata Computing Canada Inc., 25 Adelaide St East, M5C 3A1, Toronto, ON, Canada}

\author{Juan Carrasquilla}
\affiliation{Vector Institute, MaRS Centre, Toronto, Ontario, Canada M5G 1M1}
\affiliation{Department of Physics, University of Toronto, Ontario M5S 1A7, Canada}
\affiliation{Department of Physics and Astronomy, University of Waterloo, Waterloo, Ontario, Canada N2L 3G1}

\author{Alejandro Perdomo-Ortiz}
\email{alejandro@zapatacomputing.com}
\affiliation{Zapata Computing Canada Inc., 25 Adelaide St East, M5C 3A1, Toronto, ON, Canada}


\date{\today} 

\begin{abstract}
Generative modeling has seen a rising interest in both classical and quantum machine learning, and it represents a promising candidate to obtain a practical quantum advantage in the near term. In this study, we build over the framework proposed in Gili et al.~\cite{gili2022evaluating} for evaluating the generalization performance of generative models, and we establish the first quantitative comparative race towards practical quantum advantage (PQA) between classical and quantum generative models, namely Quantum Circuit Born Machines (QCBMs), Transformers (TFs), Recurrent Neural Networks (RNNs), Variational Autoencoders (VAEs), and Wasserstein Generative Adversarial Networks (WGANs). After defining four types of PQAs scenarios, we focus on what we refer to as \textit{potential PQA}, aiming to compare quantum models with the best-known classical algorithms for the task at hand.  We let the models race on a well-defined and application-relevant competition setting, where we illustrate and demonstrate our framework on 20 variables (qubits) generative modeling task. Our results suggest that QCBMs are more efficient in the data-limited regime than the other state-of-the-art classical generative models. Such a feature is highly desirable in a wide range of real-world applications where the available data is scarce.

\end{abstract}

\maketitle

\section{Introduction}\label{s:intro}

Generative modeling has become more widely popular with its remarkable success in tasks related to image generation and text synthesis, as well as machine translation~\cite{LeCun-Nature-2015, Transformer2017, Ramesh_2021, DiffusionModels2022, team2022chatgpt, ouyang2022training}, making this field a promising avenue to demonstrate the power of quantum computers and to reach the paramount milestone of \emph{practical quantum advantage (PQA)}~\cite{PerdomoOrtiz2017}. The most desirable feature in any machine learning (ML) model is \emph{generalization}, and as such, this property should be considered to assess its performance in search of PQA. However, the definition of this property in the domain of generative modeling can be cumbersome, and it is yet an unresolved question for the case of arbitrary generative tasks~\cite{alaa2021faithful}. Its definition can take on different nuances depending on the area of research, such as in computational learning theory~\cite{Vapnik99} or other practical approaches~\cite{zhao2018bias,nica2022evaluating}. Ref.~\cite{gili2022evaluating} defines an unambiguous framework for generalization on discrete search spaces for practical tasks. This approach puts all generative models on an equal footing since it is sample-based and does not require knowledge of the exact likelihood, therefore making it a model-agnostic and tractable evaluation framework. This reference also demonstrates footprints of a quantum-inspired advantage of Tensor Network Born Machines~\cite{han2018unsupervised} compared to Generative Adversarial Networks~\cite{goodfellow2016nips}. 

Remarkably, there is still a lack of a concrete quantitative comparison between quantum generative models and a broader class of classical state-of-the-art generative models, in search of PQA. In particular, quantum circuit Born machines (QCBMs)~\cite{Benedetti2019} have not been compared up-to-date with other classical generative models in terms of generalization, although they have been shown recently for their ability to generalize~\cite{Gili2022}. In this paper, we aim to bridge this gap and provide the first numerical comparison, to the best of our knowledge, between quantum and classical state-of-the-art generative models in terms of generalization. 

In this comparison, these models compete for PQA. For this `race' to be well-defined, it is essential to establish its rules first. Indeed, a clear-cut definition of PQA is not present in the relevant literature so far, especially when it comes to challenging ML applications such as generative modeling, or in general, to practical ML.

Previous works emphasize either computational quantum advantage, or settings that are not relevant from a real-world perspective, or scenarios that use data sets that give an advantage to the quantum model from the start (and also bear no relevance to a real-world setting)~\cite{Havlicek2019, boixo2018characterizing, Google2019supremacy, bouland2019complexity, Madsen2022}. One potential exception would be the case of Ref.~\cite{Huang2022}, which showed an advantage for a quantum ML model in a practical setting. However, besides the unresolved challenge of relying on quantum-loaded data, it is still unclear if it would be relevant to some concrete real-world and large-scale applications, although the authors mention some potential applications in the domain of quantum sensing.

We acknowledge as well previous works that have attempted or proposed ways to perform model comparisons, within generative models and beyond. For instance, a recent work~\cite{VarBench2023} has developed a novel metric for assessing the quality of variational calculations for different classical and quantum models on an equal footing. Another recent study~\cite{riofrio2023performance} proposes a detailed analysis that systematically compares generative models in terms of the quality of training to provide insights on the advantage of their adoption by quantum computing practitioners, although without addressing the question of generalization. In another recent work~\cite{QuantumUtility23}, the authors propose the generic notion of quantum utility, a measure for quantifying effectiveness and practicality, as an index for PQA, but this work differs from our study in the sense that PQA is defined in a broad perspective as the ability of a quantum device to be either faster, more accurate or demanding less energy compared to classical machines with similar characteristics in a certain task of interest. Others have emphasized quantum simulation as one of the prominent opportunities for PQA~\cite{daley2022practical}. In our paper, we share the long-term goal of identifying practical use cases for which quantum computing has the potential to bring an advantage. However, our work is focused on generative models and their generalization capabilities, which is the gold standard to measure the performance of generative ML models in real-world use cases. 

\begin{figure*}[htb]
\includegraphics[width=0.95\linewidth]{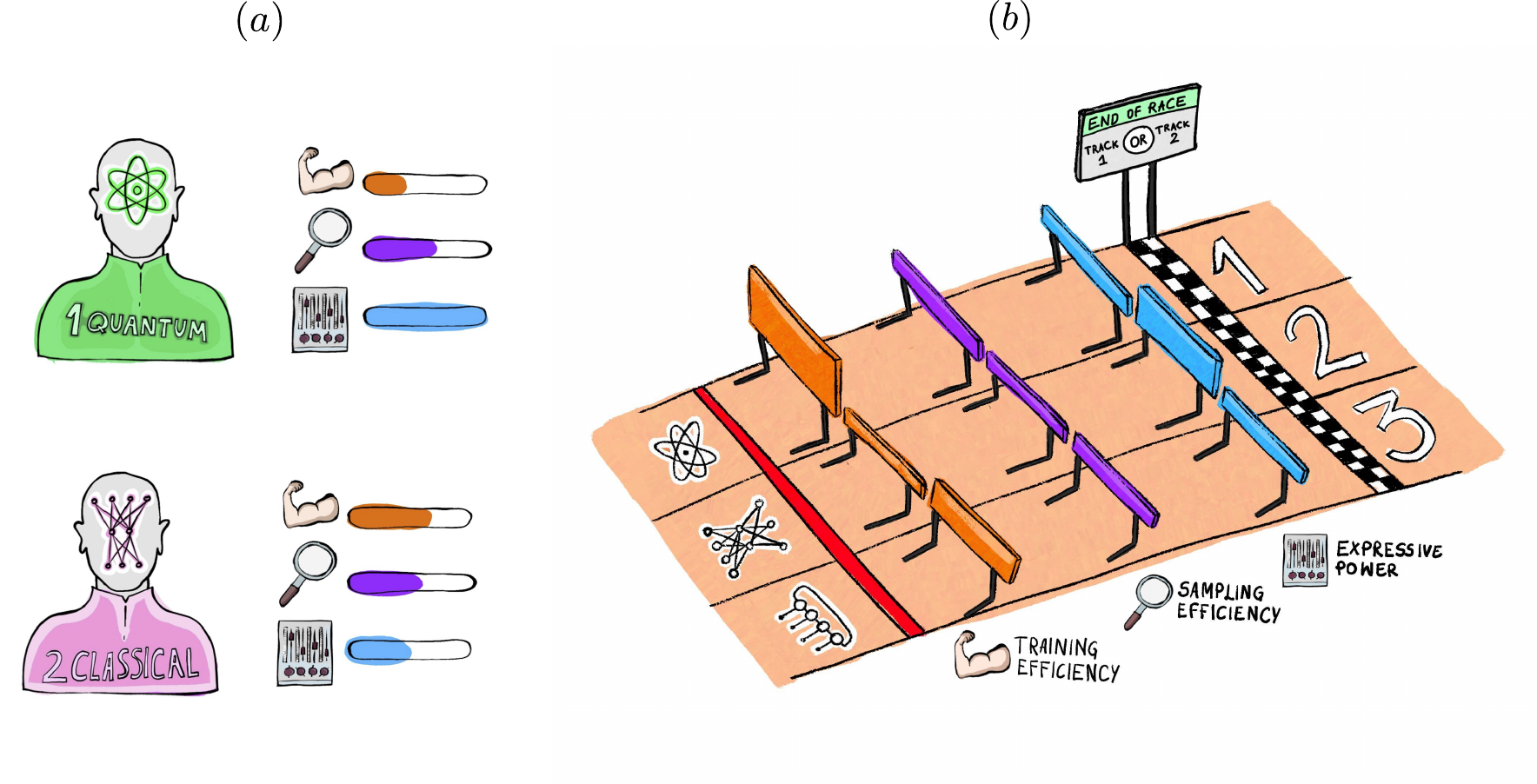}
\caption{\textbf{The practical quantum advantage (PQA) race: a sports analogy}. In panel (a), each runner (generative model) is characterized by (some of) its strengths and weaknesses, namely: training efficiency, sampling efficiency, and expressive power. Note that the power bars are indicative, and that is far from trivial to determine, but some insights can be obtained from intuition from the theoretical characterization of some of the models, e.g., via computational quantum advantage papers, or known properties or highlights for each model. A complete characterization of the runner can be used to identify the odds-on favorite, independent of the specific race context. In panel (b), the different runners are embedded into a context (i.e., `the real-world application setting') represented as a concrete instance of a hurdles race. They all run the same race, but they see the hurdles differently according to their strengths and weaknesses. The runners can compete on different tracks, for instance, on shorter or longer tracks. For the PQA race to be well defined, it is necessary to clearly state what track is taken under examination. In this study, we propose two tracks, motivated by the limitation of sampling or cost evaluation budget. Once the track is selected, we can evaluate runners using different criteria: application-driven metrics need to be defined to fully characterize the race. Our evaluation criterion is the quality-based generalization, with appropriate metrics defined in Sec.~\ref{ss:metrics} (see also Fig.~\ref{f:lenses} for further specific details).}
\label{f:sports-analogy}
\end{figure*}

In summary, the goal of this framework and of this study is to set the stage for a quantitative race between different state-of-the-art classical and quantum generative models in terms of generalization in search of PQA, uncovering the strengths and weaknesses of each model under realistic ``race conditions" (see Fig.~\ref{f:sports-analogy}). These competition rules are defined in advance before the fine-tuning of each model and dictated by the desired outcome from real-world motivated metrics and limitations, making our framework  application and/or commercially relevant from the start. Hence, we consider this formalization to be one of the main contributions of this work. This focus is motivated by the growing interest of the scientific and business community in showcasing the value of quantum strategies compared to conventional algorithms, and provides a common ground for a fair comparison based on relevant properties.

This paper is structured as follows. In Sec.~\ref{s:methods}, we provide details about the metrics we use to compare our generative models. We also contribute to the paramount yet unanswered question of better defining PQA in the scope of generalization by formalizing several types of PQA and the specific rules for the competition. In Sec.~\ref{s:results}, we show that QCBMs are competitive with the other classical state-of-the-art generative models and provide the best compromise for the requirements of the generalization framework we are adopting. Remarkably, we demonstrate that QCBMs perform well in the low-data regime, which constitutes a bottleneck for deep learning models~\cite{Deep_learning_appraisal2018, zhang2018strategy, Austin2020} and which we believe to be a promising setting for PQA.

\section{Methods}\label{s:methods}
\subsection{Generalization Metrics}\label{ss:metrics}

The evaluation of unsupervised generative models is a challenging task, especially when one aims to compare different models in terms of generalization. In this work, we focus on discrete probability distributions of bitstrings where an unambiguous definition of generalization is possible~\cite{gili2022evaluating}. Here we start from the framework provided in Ref.~\cite{gili2022evaluating} that puts different generative models on an equal footing and allows us to assess the generalization performances of each generative model from a practical perspective.

In this framework, we assume that we are given a solution space $S$ that corresponds to the set of bitstrings that satisfy a constraint or a set of constraints, such that $|S| \leq 2^{N_{\text{var}}}$ where $N_{\text{var}}$ is the number of binary variables in a bitstring. A typical example is the portfolio optimization problem, where there is a constraint on the number of assets to be included in a portfolio. Additionally, we assume that we are given a training dataset $\mathcal{D}_{\text{train}} = \{ \bm{x}^{(1)}, \bm{x}^{(2)}, \ldots, \bm{x}^{(T)} \}$, where $T = \epsilon |S|$ and $\epsilon$ is a tunable parameter that controls the size of the training dataset such that $0 < \epsilon \leq 1$.

The metrics provided in Ref.~\cite{gili2022evaluating} allow probing different features of generalization. Notably, there are three main pillars of generalization: (1) pre-generalization, (2) validity-based generalization, and (3) quality-based generalization. In the main text, we focus on quality-based generalization and provide details about pre-generalization and validity-based generalization in App.~\ref{app:generalization_metrics}.

In typical real-world applications, it is desirable to generate high-quality samples that have a low cost $c$ compared to what has been seen in the training dataset. In the quality-based generalization framework, we can define the {\it minimum value} as:
\begin{equation*}
    \text{MV} = \min_{\bm{x} \in G_{\text{sol}}} c(\bm{x}),
\end{equation*}
which corresponds to the lowest cost in a given set of unseen and valid queries $G_{\text{sol}}$, which we obtain after generating a set of queries $\mathcal{G} = \{ \bm{x}^{(1)}, \bm{x}^{(2)}, \ldots, \bm{x}^{(Q)} \}$ from a generative model of interest. In our terminology, a sample $\bm{x}$ is valid if $\bm{x} \in S$ and it is considered unseen if $\bm{x} \notin D_{\text{train}}$.

To avoid the chance effect of using the minimum, we can average over different random seeds. We can also define the {\it utility} that circumvents the use of the minimum through:
\begin{equation*}
    U = \langle c(\bm{x}) \rangle_{\bm{x} \in P_5},
\end{equation*}
where $P_5$ corresponds to the set of the $5\%$ lowest-cost samples obtained from $G_{\text{sol}}$. The averaging effect allows us to ensure that a low cost was not obtained by chance.

In quality-based generalization, it is also valuable to have a diverse set of samples that have high quality. To quantify this desirable feature, we define the {\it quality coverage} as
\begin{equation*}
    C_q = \frac{|g_{\text{sol}}(c < \min_{\bm{x} \in \mathcal{D}_{\text{train}}} c(\bm{x}))|}{Q},
\end{equation*}
where $g_{\text{sol}}(c < \min_{\bm{x} \in \mathcal{D}_{\text{train}}} c(\bm{x}))$ corresponds to the set of unique valid and unseen samples that have a lower cost compared to the minimal cost in the training data. The choice of the values of the number of queries $Q$ depends on the tracks/rules of comparison presented in Sec.~\ref{sec:PQA}.

\subsection{Defining practical quantum advantage}
\label{sec:PQA}
In this work, we refer to \emph{practical quantum advantage} (PQA) as the ability of a quantum system to perform a useful task - where `useful' can refer to a scientific, industrial, or societal use - with performance that is faster or better than what is enabled by any existing classical system~\cite{alsing2022accelerating, QuantumUtility23}. We highlight that this concept differs from the \emph{computational quantum advantage} notion (originally introduced as quantum supremacy), which refers instead to the capability of quantum machines to outperform classical computers, providing a speedup in solving a given task, which would otherwise be classically unsolvable, even using the best classical machine and algorithm~\cite{Preskill2018, boixo2018characterizing, bouland2019complexity, Huang2022}. While the latter definition is focused on provable speedup, the former practical notion aims at showing an advantage, either temporal or qualitative (or both), in the context of a real-world task. 

In computational quantum complexity, demonstrating a quantum advantage often boils down to proving a theoretical result that rules out a possibility for any classical algorithm to outperform a certain classical algorithm. PQA can be defined differently based on the practical aspects of a problem of interest and the availability of classical algorithms for the specific task at hand. Here we take inspiration from Ref.~\cite{Ronnow25072014}, to define four different types of PQA.

The first version, which we refer to as {\it provable PQA} (PrPQA) has the ultimate goal of demonstrating the superiority of a quantum algorithm with respect to the best classical algorithm, where the proof is backed up by complexity theory arguments. One example of such a milestone would be to extend the quantum supremacy experiment~\cite{Google2019supremacy, liu2021redefining} to practical and commercially relevant use cases. Most likely, this scenario will require fault-tolerant quantum devices and a key practical setting is still missing for ML use cases. Examples of this would be to show a realization of Shor's algorithm at scale, although backed to some extent by complexity theory arguments. Even in that case, it is not proven that a polynomial classical algorithm exists. To the best of our knowledge, the equivalent of Shor's algorithm in the context of real-world ML tasks, i.e., useful enough to be included in the definition of {\it provable} PQA provided above, is still missing. Since we do not yet have either the theoretical or the hardware capabilities for such an ambitious goal, we focus here on the following three classes, which might be more reachable with near- and medium-term quantum devices. We define \textit{robust PQA} (RPQA) as a practical advantage of a quantum algorithm compared to the best available classical algorithms. An RPQA can be short-lived when a better classical algorithm is potentially developed after an RPQA has been established.

However, on some occasions, there is no clear consensus about the status of the best available classical algorithm as it depends on each scientific community. To go around that, we can conduct a comparison with a state-of-the-art classical algorithm or a set of classical algorithms. If there is a quantum advantage in this case, we can refer to it as {\it potential PQA} (PPQA). Within this scenario, a genuine attempt to compare against the best-known classical algorithms has to be conducted with the possibility that a PPQA is short-lived with the development or discovery of more powerful and advanced classical algorithms. A weaker scenario corresponds to the case where we promote a classical algorithm to its quantum counterpart to investigate whether quantum effects are useful. A quantum advantage in this scenario is an example of {\it limited PQA} (LPQA). A potential case is a comparison between a restricted Boltzmann machine~\cite{RBM_Hinton} and a quantum Boltzmann machine~\cite{Amin2016}. In this study, we are pushing the search for PQA beyond the LPQA scenario to a PPQA, with the hope to include a more comprehensive list of the best available classical models in our comparison in future studies.

A significant difference between our definitions and the types of quantum speedup provided in Ref.~\cite{Ronnow25072014} is that, for the PQA case, we do not require a scaling advantage as a function of the problem size. The main reason is that industry or application-relevant problems rarely vary in size, and some of them have a very well-defined unique size. For instance, the problem of portfolio optimization is usually defined over a specific asset universe of fixed size, such as the S \& P 500 market index, which involves an optimization over $N=500$ variables. As long as we have a quantum algorithm that performs better than any of the available classical solutions at the right problem size and under the exact real-world conditions, this is already of commercial value and qualifies for PQA.

In this study, we consider different generative models and let them compete for PPQA, and for this `race' to be well-defined, it is essential to establish its rules first. When searching for any of the variants of PQA in generative modeling, we argue that generalization, as defined and equipped with quantitative metrics in Ref.~\cite{gili2022evaluating}, is an essential evaluation criterion that should be used to establish whether quantum-circuit-based or quantum-inspired models have a better performance over classical ones. A fair assessment of generative models' performance consists in measuring their ability to generate novel high-scoring solutions for a given task of interest. 

To illustrate our approach, we propose a simple sports analogy. Let us consider a hurdles race, where different runners are competing against each other. Each generative model can thus be seen as a runner in such a race. Each contender has their strengths and weaknesses, which make them see hurdles differently, and which can be quantitatively analyzed and gathered to produce a full characterization of their potential performance in the race (see Fig.~\ref{f:sports-analogy}(a)). Thus, one can aim to investigate relevant model features and determine whether they constitute a strength for the model under examination. For instance, when analyzing a quantum model, one could consider its expressive power as a strength, and its training efficiency as a weakness, and vice versa for a classical model (at least as a first-order approximation). One possibility to get a more complete intuition of this characterization is to leverage the results from the \emph{computational quantum advantage} studies on synthetic datasets.

From a full characterization of all the runners, one can establish the odds-on favorite to win the race, i.e., the fastest contender. However, hurdles races take place in a specific concrete context, for instance, with given wind and track surface conditions, which affect the competition outcome significantly (see Fig.~\ref{f:sports-analogy}(b)). The PQA approach takes this concrete context into account when evaluating the contenders, who are analyzed not only `in principle' but also embedded in a specific context. For example, the track field's length is crucial for the evaluation since different runners can perform differently if the `rules of the game' are modified. The conditions of the race affect the runners' performance, which is equivalent to say that generative models are affected by factors such as the type and size of the dataset, the ground truth distribution to be learned, etc. Each instance of a generative modeling task is unique, just as the conditions for every day of the competitions could be unique. As such, the tracks and the race conditions must be specified before the competition happens, to clarify the precise setting where the search for PQA (or, in our study, for PPQA) takes place.

Lastly, we argue that, when evaluating performance in a concrete instance of a race on a given track, the measure of success for an athlete might not necessarily be attributed to the maximum speed. For instance, what matters to win a race can be the highest speed reached throughout the race, the optimal trajectory, the best technical execution of jumps, etc. Outside the analogy, for a practical implementation of a task, other factors than the speedup are likely needed to be taken into account to judge if quantum advantage has occurred. Quality-based generalization is one of these playgrounds. Although validity-based generalization is also interesting in many use cases, we focus on quality-based generalization. The latter is particularly relevant when considering combinatorial optimization problems, as suggested by the generative enhanced optimization (GEO) framework~\cite{alcazar2021enhancing}. This reference introduces a connection between generative models and optimization and  we take inspiration from this, which is in and of itself a new perspective on a family of commercially valuable use cases for generative models beyond large language models and image generation, but that is not fully appreciated yet by the ML community. Remarkably, quality-based generalization turns out to be paramount when the generative modeling task under examination is linked to an optimization problem or whenever a suitable cost/score map can be associated with the samples. It is thus desirable to learn to generate solutions with the lowest possible cost, at least lower (i.e., of better quality) compared to the available costs in a training dataset. The utility, the minimum value, and the quality coverage have been introduced precisely to quantify this capability. However, these metrics can be computed in different ways according to the main features of a specific use case, i.e., based on the track field defining the rules of the game. In Sec.~\ref{s:rules}, we propose two distinct `track fields' that give us two different lenses, according to which we conduct a comparison of generative models toward PPQA in an optimization context that takes the resource bottlenecks of the specific use case into consideration.

\subsection{Competition details}\label{s:rules}

\begin{figure*}[htb]
\centering
  \includegraphics[width=0.8\linewidth]{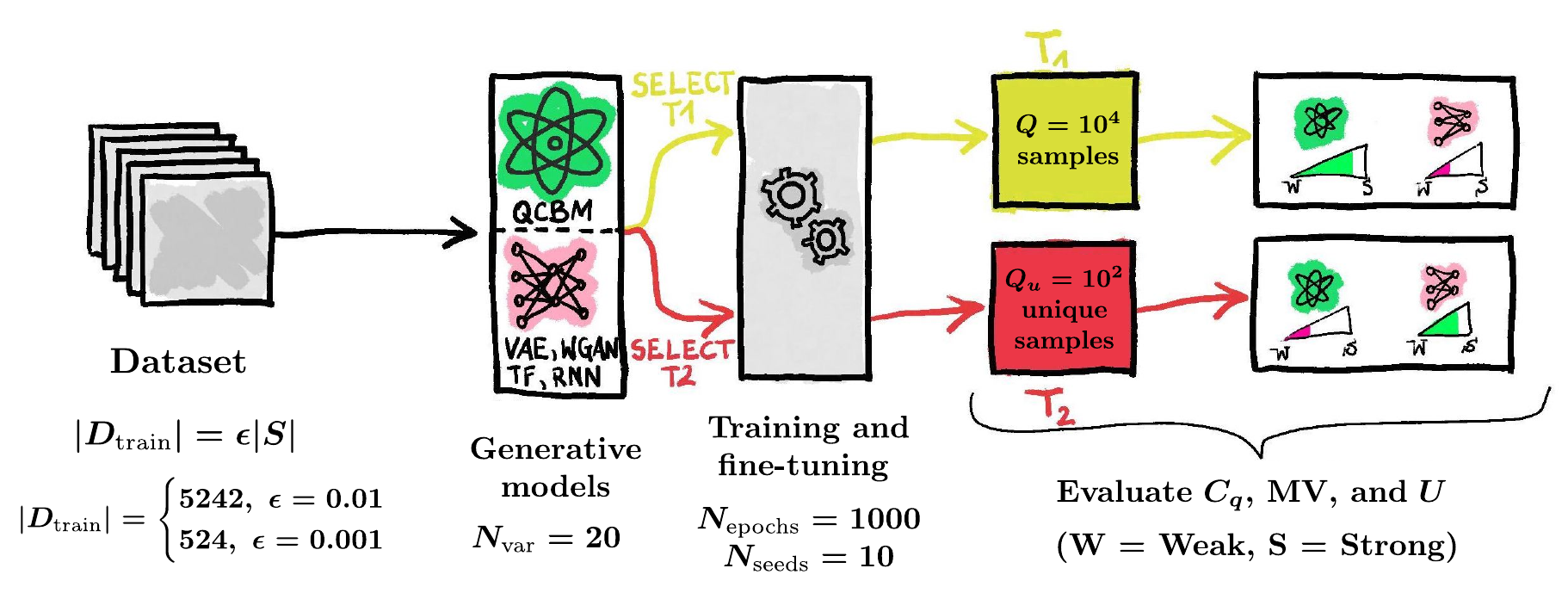}%
\caption{\textbf{An illustration of the scheme used for training and assessing the quality-based generalization of our generative models}. Given the training dataset with size $|D_{\text{train}}| = \epsilon |S|$, sampled from the \textit{Evens} distribution where $|S| = 2^{N_{\text{var}}-1}$, we choose different generative models, and select the track we want to compare them on (i.e., select the ``rules of the game" used to probe the generative models). We then train and fine-tune them using the chosen dataset. After this step, we estimate the quality-based metrics $C_q$, $MV$, and $U$ using the selected track, T1 or T2, to assess the quality of the queries generated by each model. In the first track T1, we use $Q = 10^4$ queries to estimate our metrics, whereas, for the second track T2, we require $Q_u = 10^2$ unique and valid samples at most to compute our metrics. We also choose different values of the data portion $\epsilon$ to investigate its influence on the generalization of each generative model. For a fair comparison, we use the same training budget $N_{\text{epochs}} = 1000$. Additionally, we use $N_{\text{seeds}} = 10$ different initializations for each generative model to obtain error bars on metrics.}
\label{f:lenses}
\end{figure*}

In our study, we compare several quantum and state-of-the-art classical generative models. On the quantum side, we use quantum circuit Born machines (QCBMs)~\cite{Benedetti2019} that are trained with a gradient-free optimization technique. On the classical side, we use Recurrent Neural Networks (RNN)~\cite{GRU2014}, Transformers (TF)~\cite{Transformer2017}, Variational Autoencoders (VAE)~\cite{rolfe2016discrete}, and  Wasserstein Generative Adversarial Networks (WGAN)~\cite{goodfellow2016nips}. More details about these models and their characteristics along with their hyperparameters are explained in App.~\ref{app:architectures}.

As a test bed, and to illustrate a concrete realization of our framework, we choose a re-weighted version of the \emph{Evens} (also known as parity) distribution where each bitstring with a non-zero probability has an even number of ones~\cite{Gili2022}. In this case, the size of the solution space, for $N_{\text{var}}$ binary variable, is given by $|S| = 2^{N_{\text{var}}-1}$. Furthermore, we choose a synthetic cost, called the negative separation cost $c$~\cite{Gili2022}, which is defined as the negative of the largest separation between two $1$ in a bitstring, i.e., $c(\bm{x})=-(z + 1)$, where $z$ is the largest number of consecutive zeros in the bitstring $\bm{x}$. For instance, $c(\text{`}11100011\text{'}) = -4$, $c(\text{`}10110011\text{'}) = -3$, and $c(\text{`}11111111\text{'}) = -1$. 

Given this cost function, we can define our \emph{re-weighted} training distribution $P_{\text{train}}$ over the training data, such that:
\begin{equation}
    P_{\text{train}}(\bm{x}) = \frac{\exp(-\beta c(\bm{x}))}{\sum_{\bm{y} \in D_{\text{train}}} \exp(-\beta c(\bm{y}))},
    \label{eq:softmax_train_prob}
\end{equation}
with inverse temperature $\beta \equiv \hat{\beta}/2$, where $\hat{\beta}$ is defined as the standard deviation of the scores $c$ in the training set. If a data point $\bm{x} \notin \mathcal{D}_{\text{train}}$, then we assign $P_{\text{train}}(\bm{x}) = 0$. The re-weighting procedure applied to the training data encourages our trained models to generate samples with low costs, with the hope that we sample unseen configurations that have a lower cost than the costs seen in the training set~\cite{alcazar2021enhancing}. To achieve the latter, it is crucial that the KL divergence between the generative model distribution and the training distribution does not tend to zero during the training to avoid memorizing the training data~\cite{Gili2022}. It is important to note that it is not mandatory to apply the re-weighting of the samples as part of the generative modeling task. However, the re-weighting procedure in Eq.~\ref{eq:softmax_train_prob} has been shown to help in finding high quality samples~\cite{alcazar2021enhancing,gili2022evaluating,Gili2022,Lopez-Piqueres2022}. Since all the models will be evaluated in their capabilities to generate low-cost and diverse samples, as dictated by the evaluation criteria $C_q$, $MV$, and $U$, we used the re-weighted dataset to train all the generative models studied here. In reality, the bare training set consists of $T$ data points with their respective cost values $c$ and any other transformation could be applied to facilitate the generation of high-quality samples. 

In our simulations, we choose $N_{\text{var}} = 20$ as the size of each bitstring, and we train our generative models for two training set sizes corresponding to $\epsilon = 0.001$ and $\epsilon = 0.01$ (see Fig.~\ref{f:lenses}). We choose the training data for the two different epsilons, such that we have the same minimum cost of $-12$ for the two datasets. The purpose of this constraint is to rule out the effect of the minimum seen cost in our experiments. We have selected these small epsilon values to probe the model's capabilities to successfully train and generalize in this scarce-data regime.

We focus our attention on evaluating quality-based generalization for the aforementioned generative models (the `runners') using two different competition rules (the `tracks'). These two tracks described next are motivated, respectively, by the sampling budget and the difficulty of evaluating a cost function, which are common bottlenecks affecting real-world tasks. Specifically:
\begin{itemize}
    \item Track 1 (T1): there is a fixed budget of queries $Q$ generated by the generative model to compute $C_q$, $MV$ and $U$ for the purpose of establishing the most advantageous models. This criterion is appropriate in the case where it is cheap to compute the cost associated with samples while only having access to a limited sampling budget. For instance, a definition of PPQA based on T1 can be used in the case of generative models requiring expensive resources for sampling, such as QCBMs executed on real-world quantum computers. Here, one aims to reduce the number of measurements as much as possible while still being able to see an advantage in the quality of the generated solutions.
    
    \item Track 2 (T2): there is a fixed budget $Q_u$ of unique, unseen and valid samples to compute the quality coverage, the utility and the minimum value. This approach implies the ability of sampling from the trained models repeatedly to get up to $Q_u$ unique, unseen and valid queries. Note that some models might never get to the \emph{target $Q_u$}, for instance, if they suffer from mode collapse. In this case, the metrics can be computed using the \emph{reached} $\tilde{Q}_u$. This track is motivated by a class of optimization problems where the cost evaluation is expensive. Examples of such scenarios include molecule design and drug discovery that involve clinical trials. In these settings, the cost function is expensive to compute. This track is aimed to provide a proxy reflecting these real-world use cases. In this case, one aims to avoid excessive evaluations of the cost function, i.e., for repeated samples.
\end{itemize}

Regarding the sampling budget, we use $Q = 10^4$ configurations to estimate our quality metrics for track T1. From the perspective of track T2, we sample until we obtain $Q_u = 10^2$ unique configurations that are used to compute our quality-based metrics\footnote{We checked how many samples batches are needed, and we observed that $Q = 10^4$ is enough to extract $Q_u = 10^2$ unique configurations for all the generative models in our study.}. Our metrics are averaged over $10$ random seeds for each model while keeping the same data for each portion $\epsilon$. For a fair comparison between the generative models, we conduct a hyperparameters grid search using Optuna~\cite{optuna_2019}, and we extract the best generative model that allows obtaining the lowest utility after $100$ training steps. Note that, in order to carry out the hyperparameters tuning process, one could also utilize $MV$, $C_q$, or any appropriate combination of the three metrics. Additionally, as a fair training budget, we train all our generative models for $N_{\text{epochs}} = 1000$ steps. We compute our quality-based generalization metrics for tracks T1 and T2 after each 100 training steps. We do not include this sampling cost in the evaluation budget ($Q$ or $Q_u$), as in this study we are not focusing on the training efficiency of these models, so we allow potentially unlimited resources for the training process. However, for a more realistic setting, the sampling budget could be customized to keep the training requirements into account. For clarity, Fig.~\ref{f:lenses} provides a schematic representation of our methods. The hyperparameters of each architecture and the parameters count are detailed in Tab.~\ref{tab:hyperparams}.

\section{Results and Discussion}~\label{s:results}
\begin{figure*}[htb]
    \centering
    \includegraphics[width =0.9\linewidth]{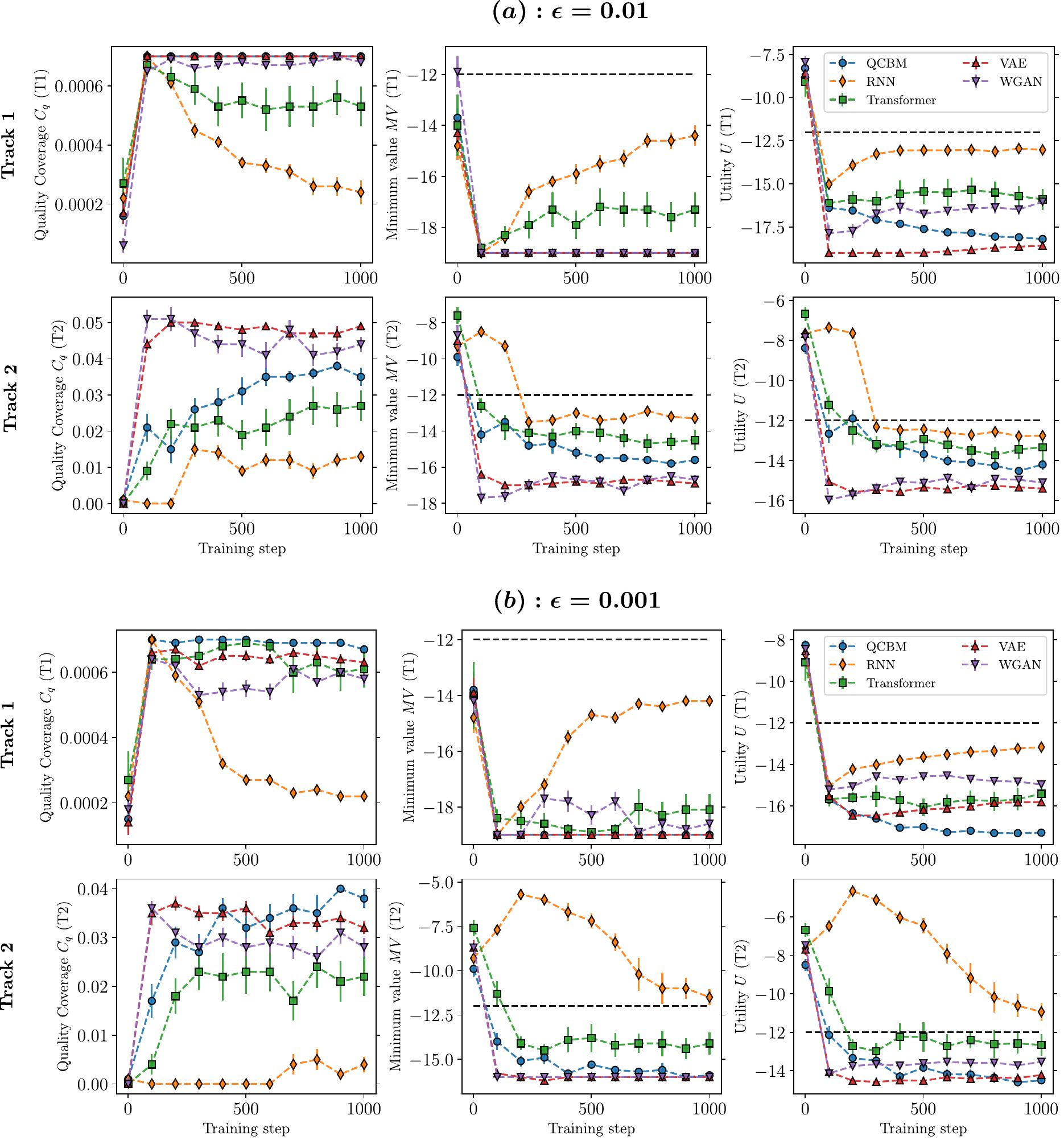}
    \caption{\textbf{A quality-based generalization comparison between QCBMs, RNNs, TFs, WGANs and VAEs.} Here, we plot the quality coverage, utility, and minimum value for the two tracks, T1 (top row) and T2 (bottom row) for $N_{\text{var}} = 20$ binary variable. Additionally, the models are trained using $N_{\text{seeds}} = 10$ random seeds, and the outcomes of the metrics are averaged over these seeds with error bars estimated as one standard deviation, which can be computed for each metric as $\sqrt{\text{Variance}/N_{\text{seeds}}}$. Panel (a) corresponds to $\epsilon = 0.01$, hence to a size of the training dataset of $5242$. Here the VAE provides the best overall performance for T1 whereas the WGAN is superior compared to the other models for T2. Panel (b) corresponds to $\epsilon = 0.001$, hence to a smaller size of the training dataset equal to $524$. From the T1 point of view, we observe that the QCBM obtains the lowest utility compared to the other models while having a competitive diversity of high-quality solutions. From the perspective of T2, QCBMs are competitive with the VAE and ahead of the WGAN, the TF, and the RNN. These results highlight the efficiency of the QCBMs in the scarce-data regime. Note that the dashed horizontal lines correspond to the minimum cost of $-12$ in the training data.}
    \label{fig:quality_metrics}
\end{figure*}

\begin{figure*}[htp]
    \centering
    \includegraphics[width =0.9\linewidth]{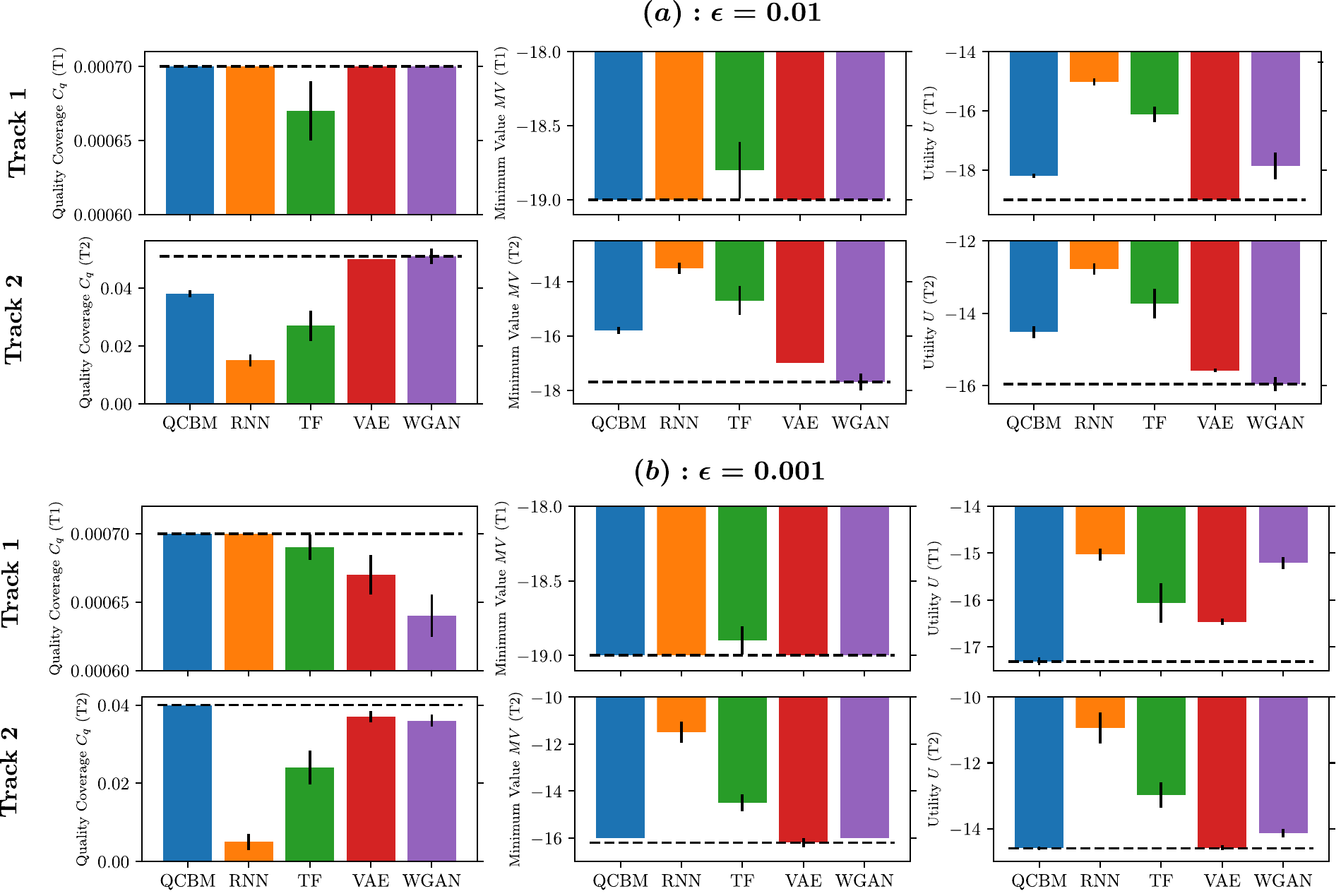}
    \caption{\textbf{Summary of the best quality-based metrics of QCBMs, RNNs, TFs, VAEs and WGANs where the setup is the same as Fig.~\ref{fig:quality_metrics}.} In panel (a), we represent the best quality metrics with $\bm{\epsilon = 0.01}$. Here we observe that the VAE has the best performance for track T1, whereas the WGAN is the best model for track T2. In panel (b), we represent our best results for $\bm{\epsilon = 0.001}$. Here we remark that the QCBM has optimal performances for track T1 and is competitive with the other models on track T2 in terms of $MV$ and $U$ while providing a better $C_q$.}
    \label{fig:best_quality_metrics}
\end{figure*}

In this section, we show the generalization results of the different generative models for the two levels of data availability, $\epsilon = 0.01, 0.001$, and for the two different tracks, T1 and T2. We start our analysis with $\epsilon = 0.01$ as illustrated in Fig.~\ref{fig:quality_metrics}(a). By looking at the first track T1, and focusing on the $MV$ results, we observe that the models experience a quick drop for the first 100 training steps. It is also interesting to see that all the models produce samples with a cost lower than the minimum cost value provided in the training set samples. Furthermore, we can see that VAEs, WGANs, and QCBMs converge to the lowest minimum value of $-19$, whereas RNNs and TFs jump to higher minimum values with more training steps. In this case, these two models gradually overfit the training data and generalize less to the low-cost sectors. This point highlights the importance of early stopping or monitoring our models during training to obtain their best performances. The utility (T1) provides a complementary picture, where we observe the VAE providing the lowest utility throughout training, followed by the QCBM and then by the other generative models. This ranking highlights the value of QCBMs compared to the other classical generative models. One interesting feature of QCBMs compared to the other models is the monotonic decrease of the utility in addition to its competitive diversity of samples, as illustrated by the quality coverage (T1). The quality coverage also shows the ability of QCBMs, in addition to VAEs and WGANs, to generate a diverse pool of unseen solutions with a lower cost compared to the costs shown in the training data. From the point of view of the second track T2, we observe that the WGAN has the best performance in terms of the three metrics. Additionally, all the models are still generalizing to configurations with a lower cost compared to what was seen in the training data. A complementary picture of the best quality metrics throughout training is provided in Fig.~\ref{fig:best_quality_metrics}(a) for clearer visibility of the ranking of generative models in our race.

We now focus our attention on the results obtained for the degree of data availability corresponding to $\epsilon = 0.001$ as illustrated in Fig.~\ref{fig:quality_metrics}(b). We again observe that all the models are generalizing to unseen configurations with a lower cost than the minimum cost seen in the training data. Remarkably, for T1, we highlight that the QCBM provides the lowest utility compared to the other models while maintaining a competitive minimum value and diversity of high-quality solutions. For the second track, T2, we observe that the QCBM is competitive with the VAE while providing the best quality coverage $C_q$. This point is clearer when analyzing and comparing the best quality-based metrics values in Fig.~\ref{fig:best_quality_metrics}(b).

Overall, QCBMs provide the best quality-based generalization performances compared to the other generative models in the low-data regime with the limited sampling budget, i.e., for $\epsilon = 0.001$ and T1 with a sampling budget of $Q = 10^4$ queries. This efficiency in the low-data regime is a highly desirable feature compared to classical generative models, which are known in real-world settings to be data-hungry~\cite{Deep_learning_appraisal2018, zhang2018strategy, Austin2020}. It is worthwhile to note that the used QCBM has the lowest number of parameters compared to the other generative models as outlined in App.~\ref{app:architectures}. Although using the parameters count to compare substantially different generative models is not necessarily a well-founded method (even if widespread), we highlight that the quantum models are able to achieve results that are competitive with classical models that have significantly more parameters, sometimes one to two order(s) of magnitude more. Overall, these findings are promising steps toward identifying scenarios where quantum models can provide a potential advantage in the scarce data regime. More details about the best results obtained by our generative models can be found in App.~\ref{app:additional_data}.

Finally, we would like to note that QCBMs are also competitive with RNNs and TFs in terms of pre-generalization and validity-based generalization metrics for both data availability settings, $\epsilon = 0.001, 0.01$, as outlined in App.~\ref{app:generalization_metrics}. The VAE and the WGAN tend to sacrifice these aspects of generalization compared to quality-based generalization (see App.~\ref{app:generalization_metrics}). Remarkably, the QCBM provides the best balance between quality-based and validity-based generalization.

\section{Conclusions and Outlooks}\label{s:conclusions}
In this paper, we have established a race between classical and quantum generative models in terms of quality-based generalization and defined four types of practical quantum advantage (PQA). Here, we focus on what we referred to as \textit{potential PQA} (PPQA), which aims to compare quantum models with the best-known classical algorithms to the best of our efforts and compute capabilities for the specific task at hand. We have proposed two different competition rules for comparing different models and defining PPQA. We denote these rules as tracks based on the race analogy. We have used QCBMs, RNNs, TFs, VAEs and WGANs to provide a first instance of this comparison on the two tracks. The first track (T1) relies on assuming a fixed sampling budget at the evaluation stage while allowing for an arbitrary number of cost function evaluations. In contrast, the second track (T2) assumes we only have access to a limited number of cost function evaluations, which is the case for applications where the cost estimation is expensive. We also study the impact of the degree of data available to the models for their training. Our results have demonstrated that QCBMs are the most efficient in the scarce-data regime and, in particular, in T2. In general, QCBMs showcase a competitive diversity of solutions compared to the other state-of-the-art generative model in all the tracks and datasets considered here.

It is important to note that the two tracks we chose for this study are not comprehensive, even though they are well motivated by plausible real-world scenarios. One could also use different rules of the game where, for example, the training data can be updated for each training step, as it is customary in the generator-enhanced optimization (GEO) framework~\cite{alcazar2021enhancing}, or where the overall budget takes into account the number of samples required during training. The two tracks introduced here serve the purpose of illustrating the possibilities ahead from this formal approach, In particular, such an approach helps to unambiguously specify the criteria for establishing PQA for generative models in real-world use cases, especially in the context of generative modeling with the goal of generating diverse and valuable solutions, which could boost in turn the solution to combinatorial optimization problems. This characterization is a long-sought-after milestone by many application scientists in the quantum information community, and we believe this framework can provide valuable insights when analyzing the suitability of the adoption of quantum or quantum-inspired models against state-of-the-art classical ones.

Despite the encouraging results obtained from our quantum-based models, we foresee a significant space for potential improvements regarding all the generative models used in this study and some not explored here. In particular, one can embed constraints into generative models such as in $U(1)$-symmetric tensor networks~\cite{Lopez-Piqueres2022} and $U(1)$-symmetric RNNs~\cite{Hibat_Allah_2020, morawetz2021u}. Furthermore, including other state-of-the-art generative models with different variations is vital for establishing a more comprehensive comparison. Lastly, the extension of this work to more realistic datasets is also crucial in the quest to investigate generalization-based PQA. We hope that our work will encourage more comparisons with a broader class of generative models and that it will be diversified to include more criteria for comparison into account.\\

\begin{acknowledgments} 

We would like to thank Brian Chen for his generous comments and suggestions which were very helpful. We also acknowledge Javier Lopez-Piqueres, Daniel Varoli, Vladimir Vargas-Calderón, Brian Dellabetta and Manuel Rudolph for insightful discussions. We also acknowledge Zofia Włoczewska for assistance in designing our figures. Our numerical simulations were performed using Orquestra\textsuperscript{TM}. M.H acknowledges support from Mitacs through Mitacs Accelerate. J.C. acknowledges support from Natural Sciences and Engineering Research Council of Canada (NSERC), the Shared Hierarchical Academic Research Computing Network (SHARCNET), Compute Canada, and the Canadian Institute for Advanced Research (CIFAR) AI chair program.

\end{acknowledgments}

\bibliography{main}

\clearpage
\newpage

\appendix

\section{Generative models}\label{app:architectures}

In this appendix, we briefly introduce the generative models used in this study. We also summarize their hyperparameters in Tab.~\ref{tab:hyperparams}.

\subsection{Quantum Circuit Born Machines (QCBMs)}

Quantum circuit Born machines (QCBMs) are a class of expressive quantum generative models based on parametrized quantum circuits (PQCs)~\cite{Benedetti2019}. Starting from a fixed initial qubit configuration $\ket{\bm{0}}^{\otimes N}$, we apply a unitary $U(\bm{\theta})$ on top of this state. Projective measurements are performed at the end of the circuit to get configurations that are sampled, according to Born's rule, from the squared modulus of the PQC's wave function. One can use different topologies that describe the qubit connectivity to model the unitary $U(\bm{\theta})$. Our study uses the line topology, where each qubit is connected to its nearest neighbors in a 1D chain configuration~\cite{Benedetti2019,benedetti2019parameterized}. More details can be found in Ref.~\cite{Gili2022}. 

To train our QCBM, we compute the KL divergence between the softmax training distribution $P_{\text{train}}$ (see Eq.~\eqref{eq:softmax_train_prob}) and the QCBM probability distribution $P_{\text{QCBM}}$ as 
\begin{equation*}
    \text{KL}(P_{\text{train}} || P_{\text{QCBM}}) = \sum_{\bm{x}} P_{\text{train}}(\bm{x}) \log \left( \frac{P_{\text{train}}(\bm{x})}{\max(\delta,P_{\text{QCBM}}(\bm{x}))} \right),
\end{equation*}
where $\delta = 10^{-8}$ is a regularization factor to avoid numerical instabilities. The optimization of the parameters $\bm{\theta}$ is performed using a gradient-free method called CMA-ES optimizer~\cite{hansen2016cma, hansen2019pycma} after randomly initializing the parameters of the PQC. In our simulations, we used $8$ layers, which provided the best utility $U$ compared to $2, 4$ and $6$ layers with line topology.
For the QCBM and the following classical generative models, the optimal hyperparameters are obtained with a grid search with the condition of getting the lowest utility after $100$ training iterations.

We have also performed simulations with the all-to-all topology~\cite{Benedetti2019}; however, they lead to sub-optimal performances compared to the line topology for trainability reasons at the scale of our experiments. In Ref.~\cite{Gili2022}, we did not observe this issue for an all-to-all topology at the scale of $12$ qubits. To further improve the trainability of QCBMs, there is a potential for using pre-training of QCBMs with tensor networks as suggested in Ref.~\cite{rudolph2022synergistic}. As a final note, since the exact computation of $P_{\text{QCBM}}$ is not tractable for large system sizes, one could consider the use of sample-based cost functions such as Maximum Mean Discrepancy (MMD) for a large number of qubits~\cite{liu2018differentiable}. Extending this study to an experimental demonstration on currently available quantum devices is also crucial to evaluate the impact of noise on the model's performance.

\subsection{Recurrent Neural Networks (RNNs)}

Recurrent Neural Networks (RNNs) are unique architectures traditionally used for applications in natural language processing such as machine translation and speech recognition~\cite{lipton2015RNN, Bengio-Book}. They are known for their ability to simulate Turing machines~\cite{RNNTuring} and for their capability of being universal approximators~\cite{Shafer2006}. RNNs take advantage of the probability chain rule to generate uncorrelated samples autoregressively as follows:
\begin{align}
    P&_{\rm RNN}(\sigma_1, \sigma_2, \ldots, \sigma_N) = \nonumber\\
    &P_{\rm RNN}(\sigma_1) P_{\rm RNN}(\sigma_2 | \sigma_1) \ldots P_{\rm RNN}(\sigma_N | \sigma_1, \ldots \sigma_{N-1}).
    \label{eq:chain_rule}
\end{align}
Here $(\sigma_1, \sigma_2, \ldots, \sigma_N)$ is a configuration of $N$ bits where $\sigma_i = 0, 1$. Each conditional $P_{\rm RNN}(\sigma_i | \sigma_1, \ldots \sigma_{i-1})$ is computed using a Softmax layer as:
\begin{equation}
     P_{\rm RNN}(\sigma_i | \sigma_{<i}) = \text{Softmax} \left ( U \bm{h}_i + \bm{c} \right ) \cdot \bm{\sigma}_i.
    \label{eq:softmax_layer}
\end{equation}
$\bm{\sigma}_i$ is a one-hot encoding of $\sigma_i$ and `$\cdot$' is the dot product operation. The weights $U$ and the biases $\bm{c}$ are the parameters of this Softmax layer. The vector $\bm{h}_i$ is called the memory state with a tunable size $d_h$, which we call the hidden dimension. For the simplest (vanilla) RNN, $\bm{h}_i$ is computed using the following recursion relation:
\begin{equation}
    \bm{h}_i = f(W \bm{h}_{i-1} + V \bm{\sigma}_{i-1} + \bm{b}),
\end{equation}
where $W$, $V$ and $\bm{b}$ are trainable parameters and $f$ is a non-linear activation function. Furthermore, $\bm{\sigma}_{0}, \bm{h}_0$ are initialized to zero. Typically, vanilla RNNs suffer from the vanishing gradient problem, which makes their training a challenging task~\cite{Pascanu2012}. To mitigate this limitation, more advanced versions of RNN cells have been devised, namely, the Gated Recurrent Unit (GRU)~\cite{GRU2014, lipton2015RNN}. In our experiments, we use the PyTorch implementation of the GRU unit cell~\cite{NEURIPS2019_9015}.

Similarly to the QCBM, we minimize the KL divergence between the RNN distribution~\eqref{eq:chain_rule} and the training distribution~\eqref{eq:softmax_train_prob} without a regularization factor $\delta$. For the optimization, we use Adam optimizer~\cite{Kingma2014}. To find the best hyperparameters that provided the lowest utility, we have conducted a grid search with the learning rates $\eta = 10^{-4}, 10^{-3}, 10^{-2}$ and the hidden dimensions $d_h = 8,16,32,64,128$. We find that $\eta = 10^{-3}$ and $d_h = 32$ are optimal for both the two values of the training data portion $\epsilon$.

\subsection{Transformers (TFs)}

Transformers (TFs) are attention models that have sparked exciting advances in natural language processing~\cite{Transformer2017}, namely in applications related to language translation, text summarization, and language modeling. Similarly to the RNN, TFs can have the autoregressive property, and they have been proposed in the literature as a solution to the vanishing gradient problem in RNNs~\cite{Pascanu2012}. The attention mechanism in TFs allows handling long-range correlations compared to traditional RNNs~\cite{Transformer2017, Huitao2019}. TFs can also process long data sequences compared to traditional neural network architectures.

In our simulations, we use the traditional implementation of TF decoders in Ref.~\cite{Transformer2017} without using a TF encoder in a similar fashion to the RNN. First, we embed our inputs using a multilayer perceptron (MLP) with a Leaky ReLU activation, and then we add positional encoding. Next, we use a one-layered TF with one attention head. Additionally, the outputs are passed to a two-layer feed-forward neural network with a ReLU activation in the hidden layer. To reduce the search space, the size of the hidden dimension in the feed-forward neural network is chosen to be the same as the size of the embedding dimension, which we also denote as $d_h$~\cite{Transformer2017}. The latter is fine-tuned in the range of possibilities $d_h = 8, 16, 32, 64, 128$ along with a range of learning rates $\eta = 10^{-4}, 10^{-3}, 10^{-2}$. For the training, we minimize the KL divergence between the TF probability distribution and the re-weighted training distribution~\eqref{eq:softmax_train_prob}. We find that $d_h = 64$ and $\eta = 10^{-2}$ provided the lowest utilities for both values of the training data portion $\epsilon$.

\subsection{Variational Autoencoders (VAEs)}

Variational autoencoders (VAEs) are approximate likelihood generative models that can learn to represent and reconstruct data~\cite{goodfellow2016nips}. Initially, a VAE encoder maps an input data point to a latent representation $\bm{z}$. Next, a VAE decoder uses the latent representation to reconstruct the original input data point. Finally, the VAE compares the reconstruction and the initial input data point and is trained until the reconstruction error is as small as  possible. In this case, the VAE cost function corresponds to the evidence lower bound (ELBO) of the negative log-likelihood~\cite{CyclicalVAE2019}.

In our study, the encoder and the decoder are built as an MLP with two hidden layers with a CELU ($\alpha = 2$) activation~\cite{CELU2017}. The encoder is followed by two separate MLP layers for the purpose of implementing the reparametrization trick~\cite{kingma2013auto,goodfellow2016nips}. To reduce the search space of hyperparameters, the size of these hidden layers is taken to be the same as the size of the latent representation. The latter is chosen from the range $8, 16, 32, 64, 128$. For the optimization, we explore different learning rates $\eta = 10^{-4}, 10^{-3}, 10^{-2}$. We also note that the temperature $\beta$ was not added to the training of VAEs~\cite{higgins2017betavae}. After conducting a grid search, we find that a latent dimension of $128$ and a learning rate $\eta = 10^{-3}$ are the optimal hyperparameters.

\subsection{Generative Adversarial Networks (GANs)}

Generative adversarial networks (GANs) are a class of neural networks that consists of a generator and a discriminator~\cite{goodfellow2014generative, goodfellow2016nips}. The generator is optimized based on a dataset, while the discriminator is trained to distinguish between real and generated data points. The two networks are trained together until reaching Nash equilibrium, where the generator produces data that is very similar to the original dataset while the discriminator becomes much better at distinguishing real from generated data.

In our study, we use Wasserstein GANs (WGAN)~\cite{arjovsky2017wasserstein}, which are a variant of GANs with a loss function called the Wasserstein loss (which is also known as the Earth Mover's distance). For the optimization of the loss function, we use Adam optimizer. An interesting feature about WGANs is that they are less susceptible to mode collapse. They can also be used for a wide range of applications, including audio synthesis, text generation, and image generation, and a wide range of areas of science~\cite{Dash_2021}.

In our numerical implementation, we choose the generator and discriminator as two feed-forward neural networks with two hidden layers with CELU $(\alpha = 2)$ activation~\cite{CELU2017}. For simplicity, we choose the size of the gaussian prior to being the same as the width of the hidden layers. We fine-tune the hyperparameters with a grid search on the widths $8, 16, 32, 64, 128$ and the learning rates $\eta = 10^{-4}, 10^{-3}, 10^{-2}$. We find that a prior size of $8$ and $\eta = 10^{-2}$ are optimal for the data portion $\epsilon = 0.01$, whereas a prior size of $128$ with $\eta = 10^{-3}$ works best for $\epsilon = 0.001$. 

\begin{table*}[p]
    \centering
    \footnotesize
    \begin{tabular}{|c|c|c|}\hline
       Generative model & Hyperparameter & Value \\\hline
      \multirow{4}{*}{QCBM} & Number of layers & $8$ \\
      &  Circuit topology & Line topology \\
       & Optimizer & CMA-ES optimizer with $\sigma_0 = 0.1$ \\
      & Initialization & Random initialization between $-\pi/2$ and $\pi/2$ \\
      & Total number of parameters & 256
       \\          
       \hline
      \multirow{6}{*}{RNN} & Architecture & GRU \\
      &  Number of layers & $1$ \\
       & Optimizer & Adam optimizer \\
       & Learning rate & $10^{-3}$ \\
      & Number of hidden units & $32$ \\
      & Total number of parameters & 3456
       \\    
       \hline   
      \multirow{7}{*}{TF} & Number of layers & $1$ \\
       & Number of attention heads & $1$ \\
       & Embedding dimension size & $64$ \\
      & Size of FFNN output & $64$ \\   
       & Optimizer & Adam optimizer \\
      & Learning rate & $10^{-3}$ \\
      & Total number of parameters & 25538
       \\        
       \hline   
      \multirow{12}{*}{VAE} & Prior size & $128$ \\
       & Encoder architecture & FFNN with CELU ($\alpha = 2$) activation\\
       & Encoder hidden layers width & $128$ \\
       & Number of hidden layers of encoder net & $2$ \\
       & Decoder architecture & FFNN with CELU ($\alpha = 2$) activation \\
       & Decoder hidden layers width & $128$ \\
       & Number of hidden layers of decoder net & $2$ \\
       & Temperature $1/\beta$ & $0.0$ \\
       & Optimizer & Adam optimizer \\
      & Learning rate & $10^{-3}$ \\
      & Total number of parameters & 87828
       \\      
       \hline          
      \multirow{11}{*}{WGAN $(\epsilon = 0.01)$} & Prior size & $8$ \\
       & Generator architecture & FFNN with CELU ($\alpha = 2$) activation\\
       & Generator hidden layers width & $8$ \\
       & Number of hidden layers of generator net & $2$ \\
       & Discriminator architecture & FFNN with CELU ($\alpha = 2$) activation \\
       & Discriminator hidden layers width & $8$ \\
       & Number of hidden layers of discriminator net & $2$ \\
       & Optimizer & Adam optimizer \\
       & Gradient regularization & 10.0 \\ 
      & Learning rate & $10^{-2}$ \\
      & Total number of parameters & 573
       \\
       \hline          
      \multirow{11}{*}{WGAN $(\epsilon = 0.001)$} & Prior size & $128$ \\
       & Generator architecture & FFNN with CELU ($\alpha = 2$) activation \\
       & Generator hidden layers width & $128$ \\
       & Number of hidden layers of generator net & $2$ \\
       & Discriminator architecture & FFNN with CELU ($\alpha = 2$) activation \\
       & Discriminator hidden layers width & $128$ \\
       & Number of hidden layers of discriminator net & $2$ \\
       & Optimizer & Adam optimizer \\
       & Gradient regularization & 10.0 \\
       & Learning rate & $10^{-3}$ \\
     & Total number of parameters & 54933
       \\
       \hline          
    \end{tabular}    
    \caption{Hyperparameters used to obtain the results reported in this study. FFNN stands for a feed-forward neural network. We also report the total number of parameters for each model, where the QCBM has the lowest number of variational parameters.}
    \label{tab:hyperparams}
\end{table*}

\section{Pre-generalization and validity-based generalization results}\label{app:generalization_metrics}

The focus of pre-generalization is to check whether a generative model can generate samples outside the training data. This can be done by generating a set of queries $\mathcal{G} = \{ \bm{x}^{(1)}, \bm{x}^{(2)}, \ldots, \bm{x}^{(Q)} \}$ with size $Q$ to compute the ratio:
\begin{equation*}
    E = \frac{|G_{\text{new}}|}{Q},
\end{equation*}
where $G_{\text{new}}$ is the set of unseen (i.e., out-of-training) samples among the set of generated queries $\mathcal{G}$. This metric is called {\it exploration}, and the larger its value, the more our model of interest can \emph{explore} different configurations outside the training data.

Given a set of constraints, it is also desirable that a generative model generates queries that satisfy these constraints, which are typical in combinatorial optimization problems. In this case, these queries are called valid. We can define the first validity-based metric, called the {\it generalization rate}:
\begin{equation*}
    R = \frac{|G_{\text{sol}}|}{Q},
\end{equation*}
where $G_{\text{sol}}$ is the set of valid and unseen queries. This metric quantifies the likelihood of our generated samples that are both unseen and valid, and it can be renormalized to $1$~\cite{Gili2022}. In this case, we can define the normalized rate as:
\begin{equation*}
    \tilde{R} = \frac{R}{1-\epsilon}.
\end{equation*}
Additionally, we can define the {\it generalization fidelity} as
\begin{equation*}
    F = \frac{|G_{\text{sol}}|}{|G_{\text{new}}|},
\end{equation*}
which quantifies the likelihood of a generative model to produce valid unseen samples out of the generated unseen samples, rather than invalid ones.~\footnote{This metric is not to confuse with the fidelity measure between two probability distributions or two quantum states.}.  

In some applications~\cite{nica2022evaluating}, it is also important that a generative model provides a diverse set of unseen and valid samples. For this reason, we can define a third metric called the {\it generalization coverage} defined as:
\begin{equation*}
    C = \frac{|g_{\text{sol}}|}{|S| - T} = \frac{|g_{\text{sol}}|}{|S|(1 - \epsilon)},
\end{equation*}
where $g_{\text{sol}}$ is the set of unique queries that are both unseen and valid. This metric allows quantifying the diversity of solutions as opposed to the fidelity and the rate. The coverage can also be renormalized to $1$ by defining a normalized coverage:
\begin{equation*}
    \tilde{C} = \frac{C}{1-(1-1/(|S|(1-\epsilon))^{Q}} \approx \frac{|g_{\text{sol}}|}{Q},
\end{equation*}
where the denominator corresponds to the ideal expected value of the coverage~\cite{gili2022evaluating} and the approximation holds in the regime $Q \ll |S|(1-\epsilon)$. The latter is typical when we have a relatively large number of binary variables. We highlight that it is not necessary to know $|S|$ (or $\epsilon$) in order to compute the coverage: since we are interested in using these metrics to compare models, one can simply utilize coverage ratios among models thus avoiding the need of having prior knowledge about the size of the solution space. Lastly, we specify that we only use the first track T1 to compute the validity-based metrics, since there is no cost involved in the calculation, which makes the second track T2 not applicable.

\begin{figure*}[htb]
    \centering
    \includegraphics[width = \linewidth]{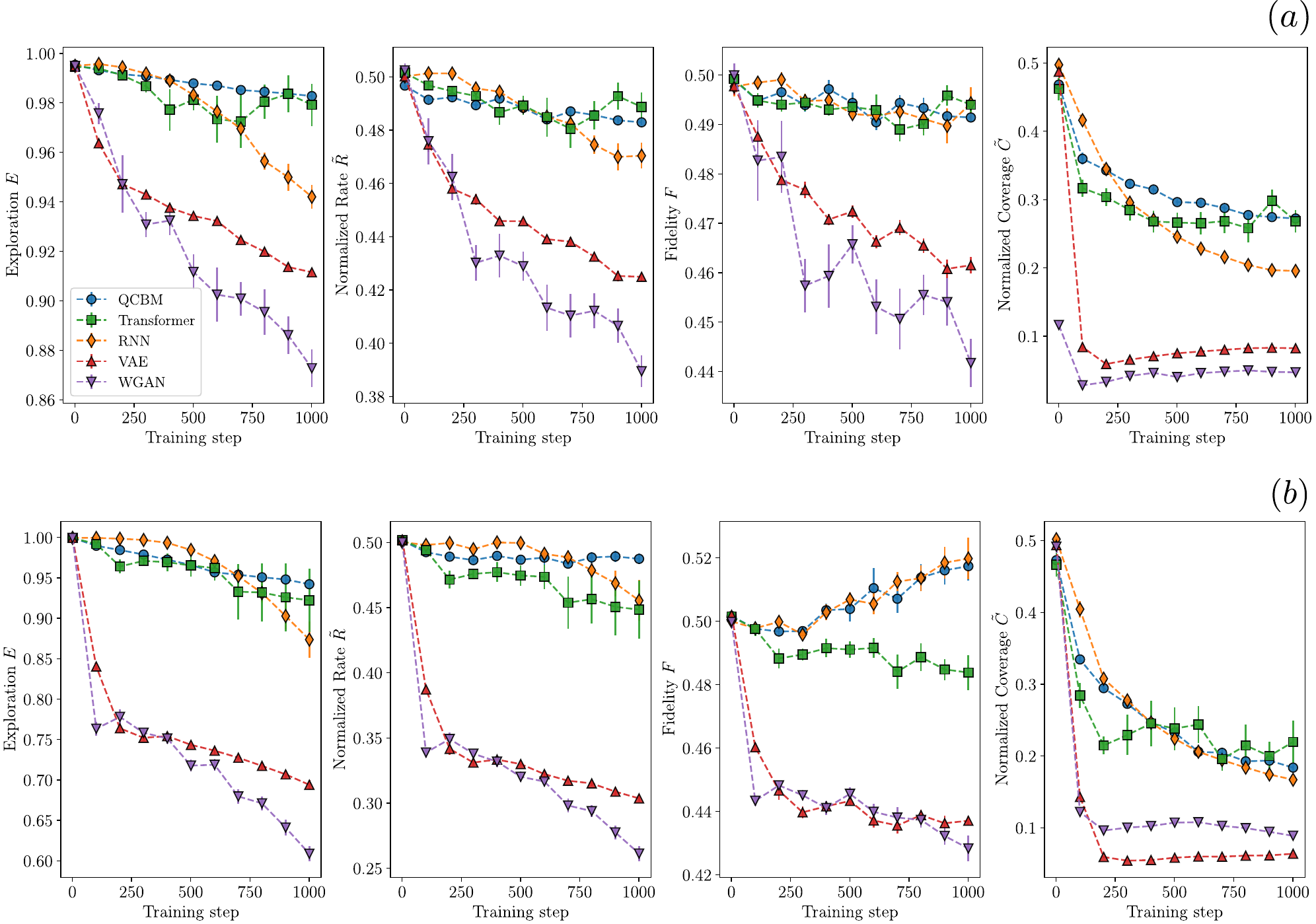}
    \caption{\textbf{Validity-based generalization comparison between QCBMs, RNNs, TFs, WGANs and VAEs} for a system size $N_{\text{var}} = 20$. Here we plot the exploration $E$, normalized rate $\tilde{R}$, fidelity $F$ and normalized coverage $\tilde{C}$ against the number of training steps for $\epsilon = 0.01$ in panel (a), and for $\epsilon = 0.001$ in panel (b). The models are trained using $N_{\text{seeds}} = 10$ random seeds, and the outcomes of the metrics are average over these seeds with errors bar estimated as the standard deviation, which can be computed for each metric as $\sqrt{\text{Var}/N_{\text{seeds}}}$.}
    \label{fig:validity_metrics}
\end{figure*}

In this section, we also present the results from the pre-generalization and the validity-based generalization metrics to provide an additional perspective on the results in the main text. In Fig.~\ref{fig:validity_metrics}, we show the exploration $E$, normalized rate $\tilde{R}$, the fidelity $F$ and the normalized coverage $\tilde{C}$ for a data portion $\epsilon = 0.01$ in panel (a) and for $\epsilon  = 0.001$ in panel (b). A common feature of the results is that QCBMs are competitive with TFs and RNNs on the validity metrics and superior to VAEs and WGANs. Besides the quality-based generalization results in the main text, these observations favor the QCBMs' ability to provide the best compromise for validity-based and quality-based generalization compared to the other generative models. The observation of VAEs being less efficient at this generalization level could be explained by the mode collapse that occurs at zero temperature. Using a finite temperature $1/\beta$ in future studies could be helpful to mitigate this limitation~\cite{higgins2017betavae}. We also observe that the WGAN has a similar behavior to the VAE, which can also be related to mode-collapse.

\section{Additional quality-based generalization results}
\label{app:additional_data}

In this appendix, we report the best values found by our generative models throughout the training shown in Fig.~\ref{fig:quality_metrics}. The error bars are computed by averaging over the outcome of $10$ different training seeds. For $\epsilon = 0.001$, we report the best performances for each metric during the training in Tab.~\ref{tab:best_result_eps1e-3}. 

\begin{table*}[]
\begin{tabular}{cccc}
\hline
{\color[HTML]{000000} \textbf{T1}}   & {\color[HTML]{000000} $\boldsymbol{C_q}$}              & {\color[HTML]{000000} \textbf{MV}}                & {\color[HTML]{000000} \textbf{U}}                  \\ \hline
{\color[HTML]{000000} QCBM} & {\color[HTML]{000000} \textbf{7e-4}}      & {\color[HTML]{000000} \textbf{-19}}      & {\color[HTML]{000000} \textbf{-17.30(8)}} \\ \hline
{\color[HTML]{000000} RNN}  & {\color[HTML]{000000} \textbf{7e-4}}      & {\color[HTML]{000000} \textbf{-19}}      & {\color[HTML]{000000} -15.03(13)}         \\ \hline
{\color[HTML]{000000} TF}   & {\color[HTML]{000000} \textbf{6.9(1)e-4}} & {\color[HTML]{000000} \textbf{-18.9(1)}} & {\color[HTML]{000000} -16.07(42)}         \\ \hline
{\color[HTML]{000000} VAE}  & {\color[HTML]{000000} 6.7(1)e-4}          & {\color[HTML]{000000} \textbf{-19}}      & {\color[HTML]{000000} -16.46(7)}          \\ \hline
{\color[HTML]{000000} WGAN} & {\color[HTML]{000000} 6.4(2)e-4}          & {\color[HTML]{000000} \textbf{-19}}      & {\color[HTML]{000000} -15.21(13)}         \\ \hline \\ \hline
{\color[HTML]{000000} \textbf{T2}}   & {\color[HTML]{000000} $\boldsymbol{C_q}$} & {\color[HTML]{000000} \textbf{MV}} & {\color[HTML]{000000} \textbf{U}} \\ \hline
{\color[HTML]{000000} QCBM} & \textbf{0.04}                & -16              & \textbf{-14.60(5)}       \\ \hline
{\color[HTML]{000000} RNN}  & 0.005(2)           & -11.5(5)        & -10.94(46)               \\ \hline
{\color[HTML]{000000} TF}   & 0.024(4)          & -14.5(4)         & 12.98(39)                \\ \hline
{\color[HTML]{000000} VAE}  & 0.037(1)                     & \textbf{-16.2(2)}         & \textbf{-14.58(7)}                \\ \hline
{\color[HTML]{000000} WGAN} & 0.036(1)                     & -16             & -14.14(12)               \\ \hline
\end{tabular}

\caption{A summary of the best values of quality coverage ($C_q$), minimum value ($MV$) and utility ($U$) for the data portion $\epsilon = 0.001$ from Fig.~\ref{fig:quality_metrics} and for the generative models (QCBM, RNN, TF, VAE and WGAN). Values in bold correspond to the best performance among the different models. Furthermore, the digits in parentheses correspond to the uncertainty over the last digit of the reported numbers. For track T1, the QCBM is the winner, whereas for track T2, QCBM is competitive with the VAE.}
\label{tab:best_result_eps1e-3}
\end{table*}

\begin{table*}[]
\begin{tabular}{cccc}
\hline
{\color[HTML]{000000} \textbf{T1}}   & {\color[HTML]{000000} $\boldsymbol{C_q}$}              & {\color[HTML]{000000} \textbf{MV}}                & {\color[HTML]{000000} \textbf{U}}                  \\ \hline
{\color[HTML]{000000} QCBM} & {\color[HTML]{000000} \textbf{7e-4}}      & {\color[HTML]{000000} \textbf{-19}}      & {\color[HTML]{000000} -18.19(7)} \\ \hline
{\color[HTML]{000000} RNN}  & {\color[HTML]{000000} \textbf{7e-4}}      & {\color[HTML]{000000} \textbf{-19}}      & {\color[HTML]{000000} -15.01(12)}         \\ \hline
{\color[HTML]{000000} TF}   & {\color[HTML]{000000} 6.7(2)e-4} & {\color[HTML]{000000} \textbf{-18.8(2)}} & {\color[HTML]{000000} -16.12(26)}         \\ \hline
{\color[HTML]{000000} VAE}  & {\color[HTML]{000000} \textbf{7e-4}}               & {\color[HTML]{000000} \textbf{-19}}      & {\color[HTML]{000000} \textbf{-19}}                \\ \hline
{\color[HTML]{000000} WGAN} & {\color[HTML]{000000} \textbf{7e-4}}               & {\color[HTML]{000000} \textbf{-19}}      & {\color[HTML]{000000} -17.86(46)}        \\ \hline \\ \hline
{\color[HTML]{000000} \textbf{T2}}   & {\color[HTML]{000000}$\boldsymbol{C_q}$} & {\color[HTML]{000000} \textbf{MV}} & {\color[HTML]{000000} \textbf{U}} \\ \hline
{\color[HTML]{000000} QCBM} & 0.038(1)            & -15.8(1)       & -14.52(16)      \\ \hline
{\color[HTML]{000000} RNN}  & 0.015(2)           & -13.5(2)         & -12.78(15)               \\ \hline
{\color[HTML]{000000} TF}   & 0.027(5)            & -14.7(5)        & -13.74(41)               \\ \hline
{\color[HTML]{000000} VAE}  & \textbf{0.05}                         & -17            & -15.58(5)                \\ \hline
{\color[HTML]{000000} WGAN} & \textbf{0.051(2)}                     & \textbf{-17.7(3)}         & \textbf{-15.96(19)}               \\ \hline
\end{tabular}
\caption{A report of the numerical values as in Tab.~\ref{tab:best_result_eps1e-3} for the data portion $\epsilon = 0.01$ reported in Fig.~\ref{fig:quality_metrics}. Values in bold correspond to the best performance among the different models. For track T1, VAE is the winner, whereas, for track T2, WGAN is superior compared to the other models.}
\label{tab:best_result_eps1e-2}
\end{table*}

\end{document}